\documentclass[prd,aps,twocolumn]{revtex4}
\usepackage{amssymb}

\usepackage{epsfig}

\begin{document}


\title{The Emergent Universe: An explicit construction}
\author{George F.R. Ellis, Jeff Murugan and Christos G. Tsagas\vskip
0.2cm} \affiliation{Department of Mathematics and Applied
Mathematics, University of Cape Town, Rondebosch 7700, Cape Town,
South Africa\vskip0.1cm}
\date{\today}

\begin{abstract}
We provide a realization of a singularity-free inflationary
universe in the form of a simple cosmological model dominated at
early times by a single minimally coupled scalar field with a
physically based potential. The universe starts asymptotically
from an initial Einstein static state, which may be large enough
to avoid the quantum gravity regime. It enters an expanding phase
that leads to inflation followed by reheating and a standard hot
Big Bang evolution. We discuss the basic characteristics of this
Emergent model and show that none is at odds with current
observations.
\end{abstract}

 \maketitle

\vskip2pc

\vskip0.5cm

\section{Introduction}
The idea that the universe we inhabit might be in a state of
eternal inflation is not new. In fact it was realized quite soon
after its initial proposal that inflation is usually future
eternal; that is, in most cases there will always be regions of
spacetime that are inflating in the future. So a natural question
arises as to whether the universe was always inflating in the
past; could inflation be {\it past eternal} also? Recent work by
several authors \cite{BV} seems to indicate that the answer to
this intriguing question is decidedly ``no"! Using
Penrose-Hawking-Geroch techniques, it is argued in detail that a
spacetime that $(i)$ is past causally simple, $(ii)$ is open,
$(iii)$ is described by Einstein's equations with a matter source
that obeys the weak energy condition, and $(iv)$ allows for
inflation to be future eternal, \textit{cannot be past null
complete}. Crucial to the argument is the assumption that the
universe is open or at least flat. As is often pointed out though,
we are emerging into an era of ``precision cosmology" and
measurements of temperature anisotropies in the Microwave
Background are able to place greater constraints on the curvature
of the universe than ever before \cite{WMAP}. The recent WMAP data seems to
point to a universe that is close to ({\it but not quite}) flat, with a
total density parameter of $\Omega_{tot}=1.02\pm 0.02$. 
Nevertheless, until we have something like a three sigma signal, it remains
premature to conclude with any certainty whether we live in a flat universe or
not
\footnote{We thank the referee for reminding us of this point.}. In
particular, the WMAP data {\it does not rule out closed models.}
In this light, it would seem that we have a promising avenue around 
the arguments
of \cite{BV}. If we live in a universe that is closed (albeit only
marginally so) today, then it was always closed, and perhaps
inflation is past-eternal after all.\\

\noindent Recently arguments were put forward for several
inflationary cosmologies that were past-eternal while avoiding any
quantum gravity regime~ \cite{EM}. Each of the proposed models is
spatially closed and described only with general relativity,
ordinary matter, and minimally coupled scalar fields. Their
existence argues against the suggestions of \cite{BV} that
inflationary universes are necessarily bounded in the past, and
shows that a quantum gravity dominated era for the universe may
not be inevitable; something that has also been noted
in~\cite{MV}. The Eddington-Lema\^{\i}tre cosmology is a
well-known example of a universe that is not past
geodesic-incomplete, because of its positively curved spatial
sections. Harrison has also given an exact solution with similar
properties~\cite{Ha}. His is a radiation-dominated closed universe
with a positive cosmological constant. It starts from an Einstein
static state, with a radius determined by the value of $\Lambda $,
before entering a never-ending period of de Sitter expansion.
However both these models do not exit inflation. Here,
following~\cite{EM}, we consider a universe filled with a
dynamical scalar field, which is past asymptotic to an Einstein
static model with a radius determined by the field's kinetic
energy. This model enters a period of de Sitter inflation that
comes naturally to an end as the scalar field starts oscillating
around the minimum of the potential, before entering the standard
hot Big-Bang expansion phase. Thus these are  singularity-free
inflationary universe models\footnote{One objection that could be raised at this
point is the persistence of closed trapped surfaces in this cosmological model and
the consequent implications for the existence of a cosmological singularity.
Indeed such {\it past closed trapped surfaces} do exist but as shown in
\cite{EllisCTC} do not imply the existence of a cosmological singularity in the
model.}, by-passing the restrictions of the
singularity theorems mentioned above \footnote{ The assumptions of
the singularity theorems were later relaxed \cite{BV} and it was
pointed out that essentially the same argument holds for any
spacetime in which null geodesics do not recross after traversing
the entire universe. This is not true for the Einstein static
spacetime so the models presented here safely fall out of the
domain of the Borde-Vilenkin theorem.}. They are finely tuned in
terms of the initial conditions, although one can use entropy
arguments to favor an initial Einstein Static phase for our
universe~\cite{G}. We consider possible implications of this
fine-tuning in the Conclusion.

\section{A potential Emergent Potential}
Slightly more than a decade ago it was shown {\cite{EllisMadsen}}
that by an inversion of the conventional viewpoint (beginning with
a scalar field whose self interaction is dictated by some
underlying particle physics considerations and subsequently
determining the evolution of the universe) scalar field dynamics
could be explicitly accounted for without a \textit{ slow-roll
approximation}. Indeed a scalar field potential $V(\phi)$ could
quite easily be `reverse engineered' for almost any desired
behavior of the scale factor $a(t)$. Several examples were
explicitly computed and are summarized:

\begin{center}
 \begin{tabular}{|l||l|l|}
 \hline
 \hspace{6mm}$a(t)$ & \hspace{23mm}$V(\phi )$ &  \\[0.5mm]\hline
 \hspace{1.5mm}$A\exp (\omega t)$ & \hspace{10mm}$3\kappa ^{-1}\omega
 ^{2}+\omega ^{2}(\phi -\phi _{0})^{2}$ &  \\[1.5mm]
 \hspace{1mm}$A\sinh (\omega t)$ & \hspace{1mm}$3\kappa ^{-1}\omega
 ^{2}+B^{2}\sinh ^{2}{\Bigl(}2\omega (\phi -\phi _{0})/B{\Bigr)}$ &
\\[1.5mm]
 \hspace{1mm}$A\cosh (\omega t)$ & \hspace{2mm}$3\kappa ^{-1}\omega
 ^{2}+B^{2}\sin ^{2}{\Bigl(}2\omega (\phi -\phi _{0})/B{\Bigr)}$ &
\\[1.5mm]
 \hspace{6mm}$At^{n}$ & \hspace{0.5mm}$(3n-1)B^{2}\exp {\Bigl(}\pm
2(\phi
 -\phi _{0})/B{\Bigr)}/2$ &  \\ \hline
 \end{tabular}
\end{center}

where $\kappa =8\pi G.$ The above respectively correspond to de
Sitter exponential expansion, de Sitter expansion from a
singularity, de Sitter expansion without a singularity, and
power-law expansion respectively \cite{EllisMadsen}. The
\textit{eternal emergent universe}, a nonsingular model past
asymptotic to an Einstein static universe with topology
$\mathbf{R}\times S^{3}$ and radius $a_{0}\gg L_{Pl},$ can be
treated in this way too. For a prescribed scale factor behavior
of $ a(t)\sim a_{0}+\exp(ht)$ an associated potential is not too
difficult to compute (see the appendix for details). While it
certainly produces the desired early time behavior, the potential
does not have a definitely zero minimum. This is an undesirable
feature as it points to a rather large cosmological constant,
although there are ways around it. Also, because of the limits on
the integration in the reverse-engineering construction, a
graceful exit from the inflationary regime is not always
guaranteed. Consequently we need to find a universe with a
potential that is essentially the same as that in the eternal
emergent universe at very early times but then goes to zero at
some finite value of the field. That is, we need to look for a
potential $V(\phi )$ that matches onto the reconstructed potential
(Fig.~\ref {ExactPotentialPic}) as $t\rightarrow -\infty ,$ with a
long flat plateau, but then has a vanishing minimum value.

\begin{figure}[tbp]
\begin{center}
\hspace*{-1cm}  \epsfxsize=7cm  \epsffile{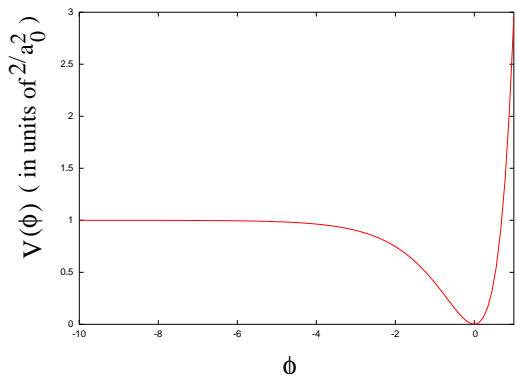}
\end{center}
\par
\caption{The Emergent potential for the parameter choice $B=1$ is
identical up to an overall rescaling to the effective potential
for the scalar field in the Einstein frame of the $R^{2}$-driven
inflation model after relabeling $\protect\phi$ as
$-\protect\varphi$.} \label{Potential}
\end{figure}

Fortunately we need not look too far. Inflationary models based on
higher derivative curvature terms go back to the remarkably
prescient work of Starobinsky~\cite{Starobinski} in which the de
Sitter phase was driven by the trace anomaly of the energy
momentum tensor (see also~\cite{HHR}). Among the variants on the
original Starobinsky model, $R^{2}$-inflation based on a
Lagrangian of the form $\mathcal{L}=R+\alpha R^{2}$ exhibits a
particularly elegant implementation of a de Sitter phase with a
linearly decaying Hubble parameter, $H$. The $R^{2}$ term in this
action is effectively an additional scalar degree of freedom which
may be absorbed by the introduction of a (non-dynamical) scalar
field ~\cite{Wands}.\footnote{ This scalar field is in fact an
auxiliary one whose `equation of motion' may be trivially solved
to show that it is nothing but the scalar curvature in disguise.}
Einstein gravity is restored by an appropriate conformal
transformation but at the expense of a \textit{dynamical} scalar
field with an interesting potential~\cite{Whitt,Maeda,KKO,Barrow}.
Since we build our model from a related potential, it is worth
seeing how this works. Starting from an $R^{2}$-modified action
\begin{eqnarray}
S &=&\int \,\mathrm{d}^{4}x\,\sqrt{-g}{\Bigl[}R+\alpha
R^{2}{\Bigr]}
\nonumber \\
&=&\int \,\mathrm{d}^{4}x\,\sqrt{-g}{\Bigl[}(1+2\alpha R)R-\alpha
R^{2}{\Bigr ]},  \label{JordonAction}
\end{eqnarray}
define $\Omega ^{2}:=1+2\alpha R$ and make the conformal
transformation $ g_{\mu \nu }\mapsto \widetilde{g}_{\mu \nu
}=\Omega ^{2}g_{\mu \nu }$ so that $\sqrt{-\widetilde{g}}=\Omega
^{4}\sqrt{-g}$ and
\begin{eqnarray}
\widetilde{R} &=&\frac{1}{\Omega ^{2}}{\Bigl[}R-6g^{\mu \nu
}\nabla _{\mu }\nabla _{\nu }(\ln \Omega )-6g^{\mu \nu }\nabla
_{\mu }(\ln \Omega )\nabla
_{\nu }(\ln \Omega ){\Bigr]}  \nonumber \\
&{}&  \label{ConformalRicci}
\end{eqnarray}
Substituting this into (\ref{JordonAction}) gives
\begin{eqnarray}
S &=&\int
\,\mathrm{d}^{4}x\sqrt{-\widetilde{g}}{\Bigl[}\widetilde{R}-\frac{
\alpha }{\Omega ^{4}}R^{2}{\Bigr]}  \nonumber \\
&+&6\int \,\mathrm{d}^{4}x\sqrt{-g}\Omega \nabla _{\mu
}{\Bigl(}g^{\mu \nu }\partial _{\nu }\Omega {\Bigr)}
\label{AlmostEinstein}
\end{eqnarray}
The last term may be evaluated by noting that the divergence
$\nabla _{\mu }X^{\mu }=(1/\sqrt{-g})\partial _{\mu
}(\sqrt{-g}X^{\mu })$. With this,
\[
\int \mathrm{d}^{4}x\sqrt{-g}\Omega \nabla _{\mu }{\Bigl(}g^{\mu
\nu }\partial _{\nu }\Omega {\Bigr)}=-\int
\mathrm{d}^{4}x\sqrt{-\widetilde{g}} \frac{1}{\Omega
^{2}}\widetilde{g}^{\mu \nu }\partial _{\mu }\Omega
\partial
_{\nu }\Omega
\]
after a boundary term is discarded. The action for the $R^{2}$
model in the Einstein frame is then written as
\begin{eqnarray}
S &=&\int
\mathrm{d}^{4}x\sqrt{-\widetilde{g}}{\Bigl\{}\widetilde{R}-\frac{
6\alpha ^{2}}{(1+2\alpha R)^{2}}{\Bigl[}\widetilde{g}^{\mu \nu
}\partial
_{\mu }R\partial _{\nu }R+  \nonumber \\
&+&\frac{1}{6\alpha }R^{2}{\Bigr]}{\Bigr\}} \label{EinsteinAction}
\end{eqnarray}
To make contact with a canonical form for scalar field
actions~\cite{Barrow}, it is usual to rewrite the scalar degree of
freedom offered by the scalar curvature as $\varphi :=\sqrt{3}\ln
(1+2\alpha R)$ so that
\[
S=\int
\mathrm{d}^{4}x\sqrt{-\widetilde{g}}{\Bigl\{}\widetilde{R}-\frac{1}{2}
\widetilde{g}^{\mu \nu }\partial _{\mu }\varphi \partial _{\nu
}\varphi - \frac{1}{4\alpha }(e^{-\varphi
/\sqrt{3}}-1)^{2}{\Bigr\}}.
\]
The effective potential in (\ref{EinsteinAction}) is just the
reflection of that in Fig.~\ref{Potential} about $\phi =0$. Note
that the parameters describing the potential are fairly rigidly
constrained (with respect to $\alpha $) by the conformal
transformation \footnote{In fact, such a potential is not unique to higher order
gravity theories. In particular, it has been shown that such potentials appear
quite naturally in models of spontaneously broken scale invariance 
\cite{Guendel}.}.

Since it's conception the $R^{2}$ model has received a significant
amount of attention, largely as a result of the fact that the
scalar field driving the de Sitter phase arises so naturally and
is not inserted \textquotedblleft by hand" solely to provide the
inflationary dynamics. Indeed it was shown in~\cite{Maeda} that
the model quite naturally supports a transient period of inflation
followed by a FRW universe. Constraints on the coupling constant $
\alpha $ are imposed by requiring that density perturbations be of
an appropriate magnitude. Consequently,
$10^{12}M_{Pl}^{-2}\lesssim \alpha \lesssim 10^{16}M_{Pl}^{-2}$.
This in turn determines the height of the plateau of the potential
to be of the order $10^{-13}-10^{-17} \times
M_{Pl}^{4}$~\cite{MMS}. This form of the potential will form the
cornerstone of our construction of the Emergent universe.

\section{Determination of the parameters in the potential}
Emergent type universes can be realized by relaxing the
constraints on the $ R^{2}$ effective potential and considering a
spacetime filled with a minimally coupled, single scalar field
$\phi$ with a potential of the general form
\begin{equation}
V= V(\phi)= \left(A\,\mathrm{e}^{B\phi}-C\right)^2+ D\,,
\label{V}
\end{equation}
where $A$, $B$, $C$ and $D$ are constants to be determined by the
specific properties of the emergent universe. Then,
\begin{equation}
V^{\prime}(\phi)=
2AB\left(A\,\mathrm{e}^{B\phi}-C\right)\mathrm{e} ^{B\phi}\,,
\label{V'}
\end{equation}
and
\begin{equation}
V^{\prime\prime}(\phi)=
2AB^2\left(2A\mathrm{e}^{B\phi}-C\right)\mathrm{e} ^{B\phi}\,,
\label{V''}
\end{equation}
where a prime indicates a derivative with respect to $\phi$.
Therefore, the above potential has  a minimum at
$\phi_0=(1/B)\ln(C/A)$ with $ V_0=V(\phi=\phi_0)=D$. Consequently,
if we want to set the minimum of the potential at the origin of
the axes, we must choose $A=C$ and $D=0$. Note that zero minimum
for $V(\phi)$ guarantees that there is no residual cosmological
constant. Then, expression (\ref{V}) reduces to
\begin{equation}
V(\phi)=A\left(\mathrm{e}^{B\phi}-1\right)^2\,.  \label{V1}
\end{equation}
By definition, the emergent universe corresponds to a
past-asymptotic Einstein-static (ES) model~\cite{EM}. This means
that
\begin{equation}
V(\phi\rightarrow-\infty)= \frac{2}{\kappa a_{0}^2}\,,
\label{V-ES}
\end{equation}
where $a_{0}$ is the radius of the initial static
model~\cite{BEMT}. It is then clear that for $a_{0}\gg L_{Pl}$,
where $L_{Pl}$ is the Planck length, the model can avoid the
quantum regime. To determine an additional parameter recall that
an ES universe filled with a single scalar field satisfies the
condition $V(\phi)=2/\kappa a_{0}^2=\dot{\phi}$, where
$\kappa=8\pi G$~\cite {BEMT}. Hence, for an initially ES state we
require that $A=2/\kappa a_{0}^{2}$, which brings the expression
of the potential down to
\begin{equation}
V(\phi)=\frac{2}{\kappa
a_{0}^2}\left(\mathrm{e}^{B\phi}-1\right)^2\,. \label{Vem}
\end{equation}
Therefore, the basic properties of the emergent universe have
already fixed three of the four parameters in the original
potential given by Eq.~(\ref{V}). The remaining parameter ($B$)
will be determined by considering other aspects of the model and
in particular by looking into the density perturbation spectrum
(see Sec.~VIII below). Nevertheless, we can still determine, the
sign of $B$ by demanding, without loss of generality, that
\begin{equation}
V^{\prime}=\frac{4B}{\kappa
a_{0}^2}\left(\mathrm{e}^{B\phi}-1\right)\mathrm{ e}^{B\phi}<0
\label{V'em}
\end{equation}
for $-\infty<\phi<0$. This means that $\mathrm{e}^{B\phi}-1<0$ and
consequently that $B>0$. Incidentally, on comparison with the
corresponding potential of the $R^{2}$ action, this is seen to
correspond to a choice of negative coupling constant $\alpha$ as
required to avoid manifesting tachyons and singular perturbative
behavior in the model~\cite{Barrow}.

\begin{figure}[tbp]
\begin{center}
\hspace*{-1cm}  \epsfxsize=7cm  \epsffile{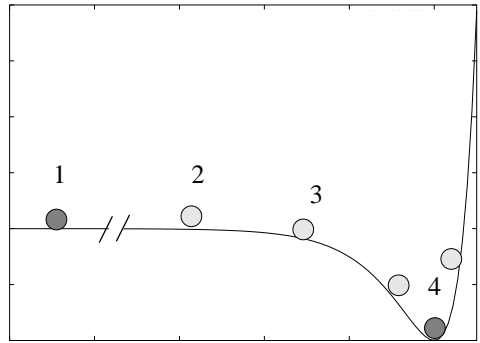}
\end{center}
\par
\caption{A schematic illustration of the scalar field evolution in
the Emergent universe. After leaving its initial static state {\bf (1)} the
model enters a slow-rolling regime (see case (ia)), or it goes
through an intermediate pre-slow-roll phase (see case(ib)) - {\bf (2)}. In
either case the scale factor grows sufficiently quickly to
mitigate neglecting the curvature effects. A period of slow-roll {\bf (3)}
inflation is followed by a re-heating phase {\bf (4)}and then by the
standard hot Big Bang evolution.} \label{Schematic}
\end{figure}

\section{Leaving the Einstein-static regime}
\noindent As the universe leaves the ES state the evolution of the
scalar field is determined by the Klein-Gordon equation
\begin{equation}
\ddot{\phi}+ 3H\dot{\phi}+ V^{\prime}=0\,,  \label{KG}
\end{equation}
where $H=\dot{a}/a$ is the Hubble parameter. Therefore, given that
$\dot{\phi }>0$, we have expansion only if
$\ddot{\phi}<-V^{\prime}$. Since the potential drops with
increasing $\phi$, we may consider the following two
alternative cases:\\

(i) $\ddot{\phi}<0$, namely a decelerated scalar field, where the
friction for the deceleration comes from the expansion. In this
case one expects the kinetic energy of $\phi$ drops. Then, since
$\dot{\phi}^2\simeq V$ initially and $V^{\prime}\simeq0$, a
slow-rolling period with $\dot{\phi}^2\ll V$
seems likely.\\

(ii) $0\leq\ddot{\phi}<-V^{\prime}$, that is a ``slowly''
accelerating scalar field. Then, one expects the kinetic energy of
the scalar field to increase. Under these conditions, a
slow-rolling regime seems unlikely. Therefore, we are left with
case (i), which splits further into two subcases:
\\

(ia) $V^{\prime}<\ddot{\phi}<0$, namely a ``weakly'' decelerated
scalar field, monitored by the familiar expression associated with
slow-roll inflation
\begin{equation}
3H\dot{\phi}+V^{\prime}=0\,,  \label{KG1b}
\end{equation}
since $|\ddot{\phi}|<|V^{\prime}|$; and\\

(ib) $\ddot{\phi}<V^{\prime}<0$, that is a ``strongly''
decelerated scalar field, described by the following form of the
Klein-Gordon equation
\begin{equation}
\ddot{\phi}+3H\dot{\phi}=0\,,  \label{KG1a}
\end{equation}
since $|\ddot{\phi}|>|V^{\prime}|$.

So, as the universe leaves the ES state, it may evolve in a number
of ways depending on the relation between $\ddot{\phi}$ and
$V^{\prime}$. Cases (i) and (ii) lead to expanding models, of
which (i) is the alternative corresponding to the Emergent
Universe. Of the two possible subcases of (i), the first leads
immediately to the standard slow-rolling inflationary regime. In
the next sections we will concentrate primarily on this particular
case. Before we proceed further, however, we should make the
following comment with regard to case (ib). It involves a strongly
decelerated, that is a very slow-rolling scalar field.
Nevertheless, Eq.~(\ref{KG1a}) also implies that
$\dot{\phi}\propto a^{-3}$. The latter suggests that, as $\phi$
drops rapidly, subcase (ib) could also lead to the familiar
slow-rolling inflation. We may test this possibility by
considering the reverse engineered potential given in the Appendix
(see Eqs.~(\ref {PhiExpression}), (\ref{VExpression})), where
$\ddot{\phi}$ dominates over $ V^{\prime}$ just like in (ib).
Then, the point where the two potentials deviate should determine
the initial conditions for the Einstein field equations that
describe subcase (ib).

\section{The duration of the slow-roll regime}
\begin{figure}[tbp]
\begin{center}
\hspace*{-1cm}  \epsfxsize=7cm  \epsffile{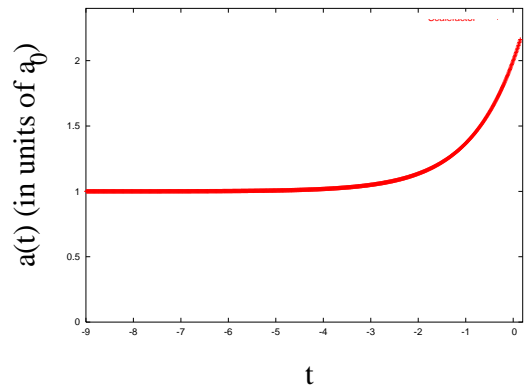}
\end{center}
\par
\caption{The initial evolution of the scale factor in the Emergent
universe.} \label{ScaleFactorPlot}
\end{figure}

We solve numerically and plot the results for the scale factor in
Fig.~\ref{ScaleFactorPlot}. The graph clearly shows that $a$ has
the familiar exponential increase associated with standard
slow-rolling inflationary models while at earlier times approaches
a constant non-zero value.

The duration of the slow-rolling inflationary regime, in a
spatially flat model, is determined by the usual slow-roll
parameters~\cite{LL}
\begin{equation}
\epsilon(\phi)={\textstyle{\frac{1}{2}}}M_{Pl}^2
\left(\frac{V^{\prime}}{V} \right)^2 \hspace{10mm} \mathrm{and}
\hspace{10mm} \eta(\phi)=M_{Pl}^2\frac{ V^{\prime\prime}}{V}\,,
\label{srpar}
\end{equation}
where $M_{Pl}$ is the Planck mass. Throughout slow-roll
approximation, the above satisfy the constraints
$\epsilon(\phi)\ll1$ and $|\eta(\phi)|\ll1$, which provide the
limits of the slow-roll regime. Applied to the potential of the
emergent universe (given by Eq.~(\ref{Vem}), with $B>0$ and for $
-\infty<\phi<0$), the above constraints read
\begin{equation}
\epsilon(\phi)=\frac{2B^2\mathrm{e}^{2B\phi}}{\left(\mathrm{e}
^{B\phi}-1\right)^2}\ll1  \label{C1}
\end{equation}
and
\begin{equation}
\left|\eta(\phi)\right|=\left|\frac{2B^2\left(2\mathrm{e}^{B\phi}-1\right)
\mathrm{e}^{B\phi}}{\left(\mathrm{e}^{B\phi}-1\right)^2}\right|\ll1\,,
\label{C2}
\end{equation}
respectively, where we have set $M_{Pl}=1$ for simplicity. It
should be emphasised that, although our model is spatially closed,
the effect of the curvature (which is dominant at very early
times) becomes negligible after few e-foldings, and only
re-emerges in the recent universe.

After a rather lengthy, but fairly straightforward analysis, one
can show that neither constraint provides a lower bound for
$\phi$. Therefore, the slow-roll regime starts at an arbitrarily
small value of $\phi$. In practice, this means that the emergent
model starts slow rolling at a few e-foldings after leaving its
initial ES state, at a finite $\phi_i$, when the curvature effects
have become negligible. Given that the static regime corresponds
to $\phi\rightarrow-\infty$, $\phi_i$ is very small and the
available number of e-foldings can be very large.

Note that, by employing (\ref{C1}) and (\ref{C2}), we can also
show that for any positive value of $B>0$ there is always a
negative value for $\phi$ at which the slow-rolling regime ends
\textit{i.e.,} where $\epsilon,\,\eta \simeq 1$. For example, when
$B=1$ we find that $\phi<-\ln(1+\sqrt{2})$ (for $M_{Pl}=1)$.

\section{The number of e-foldings}
For standard slow-roll inflation the number of e-foldings is given
by the expression~\cite{KT}
\begin{equation}
N(\phi_i\rightarrow\phi_f)=\int_{t_i}^{t_f}H\mathrm{d}t\,,
\label{N}
\end{equation}
where $\phi_i$, $\phi_f$ and $t_i$, $t_f$ are the initial and
final values of $\phi$ and $t$ respectively, and $H$ is the Hubble
parameter. The latter is generally given by~\cite{BEMT,EM}
\begin{equation}
H^2={\textstyle{\frac{1}{3}}}\kappa\left({\textstyle{\frac{1}{2}}}\dot{\phi}
+V\right)- \frac{k}{a^2}\,,  \label{Fried}
\end{equation}
where $k=0,\pm1$ is the 3-curvature index. After few e-foldings,
however, the effect of the curvature term becomes negligible,
while for slow-roll inflation we have $\dot{\phi}^2\ll V$. Thus,
throughout the slow-rolling regime $H=\sqrt{\kappa V/3}$.
Combining expression (\ref{N}) with this result we obtain
\begin{equation}
N=-\kappa\int_{\phi_i}^{\phi_f}\frac{V}{V^{\prime}}\mathrm{d}\phi\,.
\label{N2}
\end{equation}
given that $\mathrm{d}t=\mathrm{d}\phi/\dot{\phi}$ and that
$\dot{\phi} =-V^{\prime}/\sqrt{3\kappa V}$, as the slow-rolling
version of the Klein-Gordon equation (\textit{i.e.}~with
$\ddot{\phi}\simeq0$) guarantees.

\begin{figure}[tbp]
\begin{center}
\hspace*{-1cm}  \epsfxsize=7cm  \epsffile{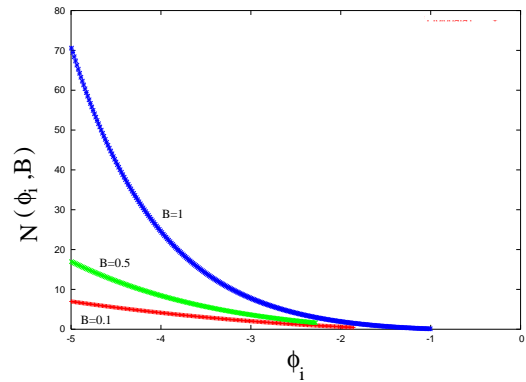}
\end{center}
\par
\caption{The number of e-folds obtained during a slowroll regime
in the Emergent model plotted against $\protect\phi_{i}$ and for
varoius values of the parameter $B$.} \label{Efolds}
\end{figure}

Applying the above to the case of the emergent universe, namely
inserting the potential (\ref{Vem}), we arrive at the expression
\begin{equation}
N=-\frac{\kappa}{2B}
\int_{\phi_i}^{\phi_f}\frac{\mathrm{e}^{B\phi}-1}{
\mathrm{e}^{B\phi}}\,,  \label{Nem}
\end{equation}
which provides the number of e-foldings associated with the
emergent universe as a function of the yet undetermined parameter
$B$. This is easily computed numerically, and in Fig.~\ref{Efolds}
we plot $N$ against the value of $\phi$ at the start of the
slowroll period for various values of the parameter $B$. Clearly,
depending on the value of $\phi_{i}$, sufficient e-folds are
easily obtainable in this model. Note that $N$, while certainly
very large, is nevertheless finite since the value of $\phi_{i}$
can be stretched back only to the point where the curvature
effects become appreciable. From a different point of view, the
reason there are only a finite number of efoldings is because the
initial value of the scale-factor is non-zero (unlike the standard
inflationary model with $K=0$).

\section{Scale factor evolution}
Starting from the Friedmann equation of the slow-rolling
inflationary regime (see Eq.~(\ref{Fried})) we have
\begin{equation}
\frac{\dot{\phi}}{a}\frac{\mathrm{d}a}{\mathrm{d}\phi}=
\sqrt{{\textstyle{ \frac{1}{3}}}\kappa V}\,.  \label{H1}
\end{equation}
At the same time, the Klein-Gordon equation gives
$\dot{\phi}=-V^{\prime}/ \sqrt{3\kappa V}$, having set
$\ddot{\phi}\simeq0$. Combining the two we arrive at
\begin{equation}
\frac{1}{a}\mathrm{d}a=-\frac{\kappa
V}{V^{\prime}}\mathrm{d}\phi\,. \label{da}
\end{equation}

Applied to the emergent universe, by using expressions (\ref{Vem})
and (\ref {V'em}), the above lead to the differential equation
\begin{equation}
\frac{1}{a}\mathrm{d}a=\frac{\kappa}{2B}\frac{1-\mathrm{e}^{B\phi}}{\mathrm{e
}^{B\phi}}\mathrm{d}\phi\,,  \label{da1}
\end{equation}
which integrated gives
\begin{equation}
\ln\left(\frac{a}{a_i}\right)=-\frac{\kappa}{2B^2}\left(\mathrm{e}^{-B\phi}-
\mathrm{e}^{-B\phi_i}\right)- \frac{\kappa}{2B}(\phi-\phi_i)\,.
\label{a}
\end{equation}
So, the value of the scale factor at the end of the slow-rolling
regime depends crucially on $\phi_i$, namely on the value of
$\phi$ at the onset of slow-roll inflation. As expected, the
smaller $\phi_i$ is the larger the final value of $a$.

\section{The density spectrum}
\begin{figure}[tbp]
\begin{center}
\hspace*{-1cm}  \epsfxsize=7cm  \epsffile{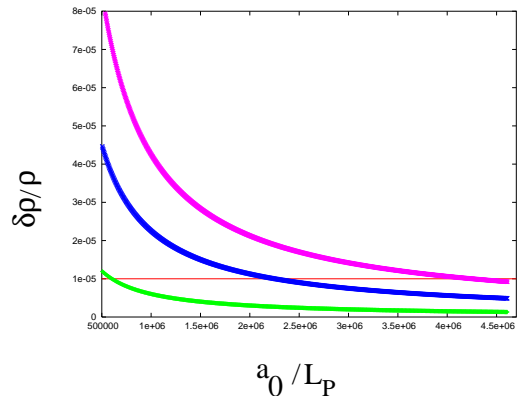}
\end{center}
\par
\caption{Plots of $\protect\delta\protect\rho/\protect\rho$
against the size of the Einstein static universe $a_{0}$ for
various values of the parameter $ B$. As $B$ increases from
$B=0.1$ through $B=1$ the value of $a_{0}$ for which
$\protect\delta\protect\rho/\protect\rho\sim 10^{-5}$ increases
from $ a_{0}\sim 6.02\times 10^{5}L_{Pl}$ to $a_{0}\sim 0.42\times
10^{7}L_{Pl}$} \label{Perturbations}
\end{figure}

For structure formation purposes it is crucial to determine the
density contrast at horizon crossing $50$ e-foldings before the
end of inflation. Following~\cite{KT} we have
\begin{equation}
\left(\frac{\delta\rho}{\rho}\right)_{Hor}\simeq
\left(\frac{H^2}{\dot{\phi}} \right)_{N=50}\simeq
-\left(\frac{3H^3}{V^{\prime}}\right)_{N=50}\,. \label{delta}
\end{equation}
During the slow-rolling regime the Hubble parameter is given by
Eq.~(\ref {Fried}) and the above takes the form
\begin{equation}
\left(\frac{\delta\rho}{\rho}\right)_{Hor}\simeq
-\sqrt{{\textstyle{\frac{1}{
3}}}\kappa^3}\left(\frac{V^{3/2}}{V^{\prime}}\right)_{N=50}\,.
\label{delta1}
\end{equation}
Substituting the potential of the emergent universe (see
Eq.~(\ref{Vem})), we obtain the following expression for the
density contrast associated with the model
\begin{equation}
\left(\frac{\delta\rho}{\rho}\right)_{Hor}\simeq
\frac{\kappa}{\sqrt{6}Ba_{0}
}\left[\frac{\left(\mathrm{e}^{B\phi}-1\right)^2}{\mathrm{e}^{B\phi}}\right]
_{N=50}\,,  \label{deltaem}
\end{equation}
which also depends on the yet undetermined parameter $B$. In view
of the COBE observations the above is constraint by
\begin{equation}
\frac{\kappa}{\sqrt{6}Ba_{0}}\left[\frac{\left(\mathrm{e}^{B\phi}-1\right)^2
}{\mathrm{e}^{B\phi}}\right]_{N=50}\simeq10^{-5}\,.  \label{COBE}
\end{equation}
Consequently, the CMB anisotropy limits allow us to express $B$ as
a function of $a_{0}$, the radius of the initial ES state, which
therefore becomes the key parameter of the emergent model. In
Fig.~\ref{Perturbations} we plot $\delta\rho/\rho$ against $a_{0}$
for several values of the parameter $B$. The appropriate value of
the Einstein radius is then read off from the intersection of the
perturbation curves with $\delta\rho/\rho = 10^{-5}$. For example,
the $B=1$ curve satisfies the density perturbation requirement
when $a_{0}\simeq10^6~L_{Pl}$. Again a comparison with the
corresponding $R^{2}$ potential shows that $\alpha\sim
a_{0}^{2}\sim 10^{12}M_{Pl}^{-2}$, within the required range for
that parameter~\cite{MMS}. This result fixes the remaining
parameter in our initial potential (see Eq.~(\ref{V})) leaving us with just one
more free parameter; the scalar field value when slow-roll commences 
$\phi_{i}$.\\

Having used the amplitude of fluctuations to
constrain the parameters of the model, it remains to check that other quantities
whose values are tightly constrained by CMB data remain within their 
experimental bounds.
Among these we count the power and polarization spectra and the spectral 
index of
the density fluctuations. As a preliminary check, we note that the spectral
index of scalar perturbations may quite easily be estimated by following the
prescription of \cite{LLMS}. We thus define the horizon-flow parameters
$\epsilon_{1}$ and $\epsilon_{2}$ through the usual slow-roll parameters
$\epsilon$ and $\eta$ as
\begin{eqnarray}
  \epsilon_{1} &:=& \epsilon/8\pi\\
  \epsilon_{2} &:=& (\epsilon - 2\eta)/4\pi
\end{eqnarray}
in terms of which the spectral index $n_{S}-1 = d\ln{\cal P}_{k}/d\ln k =
-2\epsilon_{1}-\epsilon_{2}$. Using the expressions (\ref{C1}) and (\ref{C2})
for the slowroll parameters, an estimate for the first horizon-flow parameter is
obtained as $\epsilon_{1}\sim 0.00014$. Subsequent computation of the spectral
index yields $|n_{S}-1|\sim 0.012$ or $n_{S}\sim 0.987$. This result, although
only a rough estimate is certainly compatible with those of
\cite{WMAP,LeachLiddle}. We expect that a computation of the power spectrum will
not deviate significantly from the standard result either and hope to report on
this in forthcoming work \cite{EMT2}.

With the above value for the Einstein static radius, we anticipate that the CMB 
anisotropy spectrum will have the usual Sachs-Wolfe plateau and peaks as 
confirmed by current observations~\footnote{Indeed the following observation is
relevant: the model proposed here can have the standard $S^3$
topology for the positively curved spatial surfaces, but that is
not obligatory; they could for example be Poincar\'{e}
dodecahedral spaces with positive spatial curvature, as suggested
by Luminet et al~\cite{lum}. In that case they might give a better
fit to the WMAP data at large angular scales than other
models~\cite{lum}.}. Moreover, once the scalar field potential is
fixed we have a residual {\it one-parameter family} of models - parameterized by
$\phi_{i}$ - all asymptotic to 
the same Einstein static universe and within which is contained a class of 
models with $\Omega_{tot}=1+\varepsilon$ where $\varepsilon$ is 
arbitrarily small. Since these models are all closed and begin from the same
state they serve well to illustrate the emergent model. However, the fact that
$\Omega_{tot}$ is not forced to be $1.02$ avoids both excessive fine-tuning of 
the Einstein static radius {\it and} a possible cosmic coincidence problem
\footnote{We thank the referee for pointing this out.}.

\section{Available energy for reheating}
Inflation starts at $V(\phi\rightarrow-\infty)=2/\kappa a_{0}^2$
and ends at $\phi=0$ with $V_0=0$, which means that the maximum
energy ``stored'' in the inflaton field is
\begin{equation}
V(\phi\rightarrow\infty)=\frac{2}{\kappa
a_{0}^2}\simeq\frac{10^{37} GeV^2}{ a_{0}^2}\,.  \label{Emax}
\end{equation}
In other words, the maximum energy available for reheating is
inversely proportional to the square of the radius of the initial
ES state. If $ a_{0}=10^6L_{Pl}$ we find
\begin{equation}
V(\phi\rightarrow\infty)\simeq10^{63} GeV^4\,.  \label{Emax1}
\end{equation}
Therefore, provided that thermal equilibrium has been achieved and
that the reheating process is efficient, the temperature at the
beginning of the standard Big Bang evolution can be as high as
\begin{equation}
T_{_{RH}}\simeq10^{16}GeV\,.  \label{Trh}
\end{equation}
Obviously, for less efficient reheating $T_{_{RH}}<10^{16}GeV$.

\section{Discussion}
Closed inflationary models have not had a happy time of late.
Indeed many authors have taken the latest measurements of
Microwave Background isotropies, in particular the WMAP best fit
estimate of $\Omega_{tot} = 1.02 \pm 0.02$ to signal a flat
infinite universe. Yet there are two ways to interpret these
observations. It could be that the universe {\it is} in fact flat
and more precise experiments in the future will eventually whittle
away at the error bars until $\Omega_{tot} = 1$ to within, say,
one part in a million. On the other hand, a comparison of
BOOMERANG data~\cite{Bernardis} (with a best fit of $\Omega_{tot}
= 1.02^{+0.06}_{-0.05}$) with WMAP data would seem to indicate a
convergence on $\Omega_{tot} = 1.02 > 1$ with increasing
resolution. Admittedly, the data set from which we draw this
conclusion is rather limited and it is with much eagerness that we
await results from the European Space Agency's PLANCK satellite.
However, until this debate is settled one way or the other, one is
forced to take closed models seriously. In stark contrast to some
claims in recent literature~\cite{Linde}, it is not too difficult
to construct consistent, single field inflationary models in a
closed universe. We give one such construction here (but invite
the reader to see~\cite{Lasenby-Doran} for another, simple yet
remarkably elegant model).

\noindent The ``Emergent Universe" proposed in~\cite{EM} is a
simple closed inflationary model in which the universe emerges
from an Einstein static state with radius $a_{0} \gg L_{Pl}$,
inflates and is then subsumed into a hot Big Bang era. The
attractiveness of the proposed model is that one can avoid an
initial quantum-gravity stage if the Einstein Static radius is
larger than the Planck length. One might then ask whether such a
model has a simple representation and whether it lies within the
boundaries of current observations. In this paper, we provide a
first explicit construction of such a universe. As such, it is a
{\it manifestly nonsingular} closed inflationary cosmology that
begins from a meta-stable Einstein static state and decays into a
de Sitter phase and subsequently into standard hot Big Bang
evolution. Inspired by an exact reverse-engineered potential
constructed as in~\cite{EllisMadsen} (see the appendix), this
phenomenological model employs a single scalar field with a
potential very similar to that arising in conformally transformed
$R^{2}$-inflation only with a relaxing of some of the rigidity of
that potential. In particular, beginning with a four-parameter
potential (\ref{V}), we can immediately fix two of the parameters
by fixing the origin of the potential and requiring vanishing
cosmological constant. By requiring this potential to match onto
the exact reverse-engineered one at early times ($|\phi| \gg 1$),
a third parameter of the potential is related to the radius of the
initial Einstein universe by $A^{2} = 2/\kappa a_{0}^{2}$. The
remaining two-parameter model is then shown to exhibit all the
desired properties of the Emergent universe model. Among others,
these include a sufficient number of e-folds to solve the
late-time flatness problem, a spectrum of density fluctuations
with magnitude of the order of $10^{-5}$ and sufficient energy in
the scalar field to allow for adequate post-inflationary
reheating. Of these we find that the spectrum of density
fluctuations provides another constraint that allows us to relate
the remaining two parameters so that effectively, the only
parameter in the model is the size of the Einstein static
universe. As for the re-heating energy, we show that for an
initial radius of $a_{0}=10^{6}L_{Pl}$ we can obtain re-heating
temperatures of around $10^{16}\,GeV$.\\

Of course, an immediate question arises as to a viable mechanism that
realizes the initial Einstein static universe that our model emerges
from. At this point several avenues present themselves. Among the more
promising, we note two; the first of these is the observation of
\cite{ABEN} that the Einstein static
universe is one of only two asymptotic solutions of the Ramond-Ramond
sector superstring cosmology field equations \footnote{The other being
  a linearly expanding Milne universe}. The size of the positively
curved Einstein universe in this picture is controlled by the level
number of the Kac-Moody algebra of the conformal fields living in the
compactified internal space. Indeed, this observation leads quite
naturally to the tantalizing possibility of realizing an Emergent-like
universe within a string cosmology context \cite{MEW}. A second,
equally intriguing, possibility is that the initial Einstein static
universe is created from ``nothing'' by some quantum tunneling process
\cite{Tryon1973,Vilenkin1985}. Indeed, finiteness of the tunneling action {\it
  requires} that the universe created through instanonic tunneling be
closed \cite{GLT}.
It is not implausible, then, that through spontaneous quantum
fluctuations, a closed universe could be created in a long lived but transient
Einstein static state which then makes a transition to a finite
lifetime de-Sitter and subsequent marginally closed FRW phase along
the lines described above. It remains to be seen that such a mechanism
can be concretely realized in any quantum cosmological framework.\\
   
The attractive aspects of the Einstein Static solution as a
preferred initial state for our universe have been considered in
the past. In fact, Gibbons has argued for the higher probability
of an Einstein Static initial phase based on the model's maximal
entropy~\cite{G}. Crucially, once the universe finds itself near
the static state it could remain there for an undetermined amount
of time. This is guaranteed by the neutral stability of the
Einstein Static model against inhomogeneous (either pure fluid or
pure scalar field) perturbations~\cite{har67,G,BEMT}. Typically,
expansion away from the static solution will lead to inflation
followed by the standard Hot Big Bang evolution. However, the
reader will probably by now have noted the large degree of
fine-tuning that went into setting up the initial state from which
the universe emerges. Indeed, the emergent model is a very special
trajectory in the space of possible inflationary evolutions. We
have shown existence of such models, but {\it not that they are
probable}. 

Some may regard this as a deadly blow to these models, but we
believe the case is wide open. Firstly, we note that although the
idea that the universe should be probable (and so not fine-tuned)
is the dominant paradigm in cosmology at present, there is no
scientific proof that this has to be the case. This is an unproven
and indeed unprovable philosophical assumption, which may or may
not be true \footnote{The proposal of an actually existing
multiverse, which could possibly be used to justify the idea that
the universe must indeed be probable, is also unprovable.}. It is
equally conceivable that - as was taken for granted in the past -
whatever process causes the universe to come into being prefers a
high-symmetry state. It is then relevant that the Einstein-static
model is the highest symmetry non-empty Robertson-Walker universe,
and so would be preferred by such a process.

Secondly, the models presented here show one can avoid the initial
singularity if initial conditions were fine tuned in the way
remarked on above. We believe it likely that this is a generic
result: that given the usual physics of inflationary fields in the
early universe (i.e. avoiding the introduction of `shadow matter'
which violates the weak energy condition), there is either a
singularity at the start of the universe \footnote{or at least
non-avoidance of the quantum gravity domain, when quantum gravity
processes may possibly avoid the singularity.} or a fine-tuned
initial state. This may be the real philosophical choice facing
us: to decide which is worse, a space-time singularity, with all
that that entails, or a fine tuning of initial conditions. It
certainly seems very difficult to (phenomenologically, at least)
construct a model that avoids both, and it is useful to recall
Wheeler's characterization of space time singularities caused by
gravitational collapse as the worst crisis facing theoretical
physics. Nowadays we do not perhaps take such singularities
seriously enough.

Models of the kind presented here are useful in terms of making
clear the alternatives facing us: we can indeed avoid both a
singularity and the quantum gravity regime, without introducing
any exotic physics; but there is a price to pay in terms of
fine-tuning. From some philosophical standpoints the high symmetry
of the initial state may even be an advantage.

\appendix
\section{An Exact solution}
Consider a Friedmann-Robertson-Walker (FRW) universe containing a
minimally coupled scalar field $\phi$ with Lagrangian density
\begin{eqnarray}
\mathcal{L} = \frac{1}{2}(\partial\phi)^{2} - V(\phi) =
\frac{1}{2}\dot{\phi} ^{2} - V(\phi)
\end{eqnarray}
where the last equality follows from assuming spatial homogeneity
of the $ \phi$ field \textit{i.e.} $\phi = \phi(t)$. With this
assumption, the stress-energy tensor takes the form of a perfect
fluid with energy density and pressure
\begin{eqnarray}
\rho_{\phi} &=& \frac{1}{2}\dot{\phi}^{2}+V(\phi)  \nonumber \\
p_{\phi} &=& \frac{1}{2}\dot{\phi}^{2}-V(\phi)
\label{Energy-pressure}
\end{eqnarray}
respectively \footnote{ More generally, the FRW universe may also
be assumed to contain some noninteracting perfect fluid with
energy density $\rho$ and pressure $p=w\rho $ with $-\frac{1}{3}<
w \leq 1$. The inclusion of such additional matter sources is
however unnecessary for the purpose of this work.}. The classical
equation of motion for $\phi$ that follows from variation of the
action $S = \int \mathrm{d}^{4}x\,\sqrt{-g}\mathcal{L}$ is
\begin{eqnarray}
\ddot{\phi} + 3H\dot{\phi} + \frac{dV(\phi)}{d\phi} = 0\,,
\end{eqnarray}
where the Hubble parameter $H:=\dot{a}(t)/a(t)$. The Raychaudhuri
field equation for the FRW model with scalar field matter source
and its first integral, the Friedmann equation are
\begin{eqnarray}
3\dot{H}+3H^{2} = 8\pi G{\Bigl(}V(\phi)-\dot{\phi}^{2}{\Bigr)}
\label{Ray}
\end{eqnarray}
and
\begin{eqnarray}
3H^{2}+3\frac{k}{a^{2}} = 8\pi G {\Bigl(}\frac{1}{2}\dot{\phi}^{2}
+ V(\phi){ \Bigr)}  \label{Friedmann}
\end{eqnarray}
respectively. These equations, together with (\ref{KG}) for a
closed dynamical system from which the evolution of the universe
model is determined. However, it is important to note that the
Klein-Gordon equation is \textit{auxiliary} in the sense that any
solution of (\ref{Ray}) and (\ref{Friedmann}) with nonvanishing
$\dot{\phi}$ will necessarily satisfy (\ref{KG}) so that the
dynamical system in fact contains only \textit{two} independent
equations \footnote{ If $\dot{\phi}$ does indeed vanish then the
Klein-Gordon equation is no longer a consequence of the other two
equations but may nevertheless be easily solved to yield a
constant potential $V(\phi) = const.$}. These may be combined
\cite{EllisMadsen} to give the (more convenient) equivalent set of
equations
\begin{eqnarray}
V(\phi(t)) &=& \frac{1}{8\pi G}{\Bigl(}\dot{H}+3H^{2} +
2\frac{k}{a^{2}}{
\Bigr)}  \nonumber \\
\dot{\phi}^{2}(t) &=& \frac{1}{4\pi
G}{\Bigl(}\frac{k}{a^{2}}-\dot{H}{\Bigr)} .  \label{Poteqns}
\end{eqnarray}
With these equations at hand, the potential $V(\phi)$ is
contructed by

\begin{itemize}
\item Specifying the constant $k$ and a particular (monotonic)
function $a(t) $ and  computing the associated Hubble parameter
$H$ and $\dot{H}$.

\item Checking that the constraint $k/a^{2} - \dot{H}\geq 0$ is
met. This assures the  positivity of $\dot{\phi}^{2}$ as is
necessary for a neutral scalar field.

\item Specifying an initial condition $\phi_{0}$ for $\phi(t)$ and
integrating the  second of eqs.(\ref{Poteqns}) to get $\phi(t)$.
This is then inverted (where  possible) to give $t(\phi)$.

\item Substituting into the first of eqs.(\ref{Poteqns}) to obtain
$V(t) = V(t(\phi))$  and subsequently $V(\phi)$.
\end{itemize}

Thus, as advertised, this algorithm associates to a specified
scale factor $ a(t)$ a potential $V(\phi)$ to give a model that
\textit{exactly} solves the classical inflationary field
equations. At this point some comments are in order: First, the
reconstruction of the potential is manifestly non-unique as should
be clear from the sign ambiguity in the choice of root in the $
\dot{\phi}$ equation (\ref{Poteqns}). Secondly, it is not clear
how sensitive this algorithm is to the choice of initial
conditions for $\phi$ \textit{i.e.} on where on the potential the
scalar field resides when it begins to roll. While general shape
should be independent it might be expected that the details of the
potential (location of extrema etc.) would vary with different
initial conditions. This, however, does not affect the
argument.\\

\noindent The \textit{emergent model} is essentially a modified
version of the Eddington-Lema$\hat{\mathrm{i}}$tre universe -
$k=+1$, past-asymptotically Einstein static, singularity-free,
without particle horizons and ever-inflating (see \cite{EM} for
more details). Before applying the potential reconstruction
technique to this model, a few general observations about the
scenario beg attention. The Einstein static state containing
matter with energy density $\rho_{i}$ and pressure $p_{i} =
w_{i}\rho_{i}$ ($-1/3 < w_{i} < 1$) is characterized by
\begin{eqnarray}
\frac{1}{2}(1-w_{i})\rho_{i} + V(\phi) &=& \frac{1}{4\pi G
a_{0}^{2}},
\nonumber \\
(1+w_{i})\rho_{i} + \dot{\phi}_{i}^{2} &=& \frac{1}{4\pi G},
\label{Einstein-Static}
\end{eqnarray}
where $a_{0}$ is the radius of the $S^{3}$ spatial sections of the
universe. If the sole content of this universe is the scalar
field, as is the case in this note, $\rho_{i} = 0$ and the second
of eq.(\ref{Einstein-Static}) requires that the scalar field have
non-vanishing (but constant) kinetic energy. In this model with no
matter initially present, the mechanism envisaged in \cite{EM} has
$\phi$ rolling at a constant speed along a flat potential $(V =
V_{i})$ from $\phi = -\infty$ at $t=-\infty$ to $\phi=0$ at $ t=0$
where the potential first rises (and inflation is initiated) and
then drops to a minimum at $\phi = \phi_{f}$ where the value of
the potential is $ V_{f} = \Lambda/8\pi G \ll V_{i}$. The field is
carried over the hill by its non-zero kinetic energy and slow
rolls toward the minimum where its damped oscillations reheats the
universe. What follows is an attempt to realise
this scenario.\\
\begin{figure}[tbp]
\begin{center}
\vspace{0.8cm} \hspace*{-1cm} \epsfxsize=7cm
\epsffile{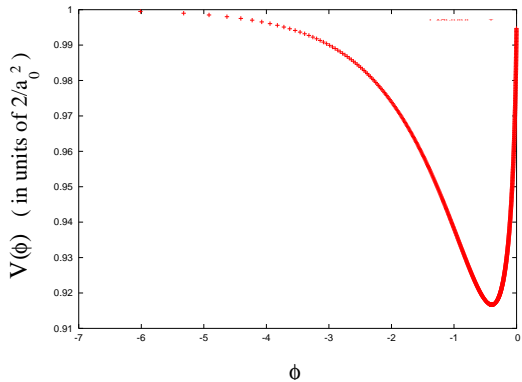}
\end{center}
\caption{A parametric plot of the re-constructed potential
corresponding to the \textit{exact solution} $a(t) = a_{0}+
\exp(ht)$ normalized with respect to the size of the Einstein
static universe. This re-constructed potential exhibits the same
shape as the effective potential of the $R^{2}$ model in the
Einstein frame but is plagued by a rather shallow minimum (only
about $ 91\%$ of its asymptotic value.)} \label{ExactPotentialPic}
\end{figure}
\noindent To this end, and to facilitate numerical computations it
will prove useful to rescale quantities of interest as
\begin{eqnarray}
V(\phi)\rightarrow M_{Pl}^{4}V(\phi)&,&\quad \phi(t)\rightarrow
M_{Pl}\phi(t),  \nonumber \\
t \rightarrow t/M_{Pl}&,&\quad a(t)\rightarrow a(t)/M_{Pl},
\label{Rescaling}
\end{eqnarray}
where $M_{Pl}^{2}:=8\pi G$. In these units the Raychaudhuri
equation and its first integral, the Friedmann equation become
\begin{eqnarray}
3\dot{H}+3H^{2} &=& V(\phi)-\dot{\phi}^{2},  \nonumber \\
3H^{2}+3\frac{1}{a^{2}} &=& \frac{1}{2}\dot{\phi}^{2} + V(\phi),
\label{NaturalFRW}
\end{eqnarray}
which, together with the Klein-Gordon equation form the dynamical
system governing the evolution of the closed scalar field
dominated FRW universe. The corresponding equations
(\ref{Poteqns}) for the potential and $\dot{\phi} (t)$ that are
the starting point for the reconstruction of the potential take
the form
\begin{eqnarray}
V(\phi(t)) &=& \dot{H}+3H^{2} + \frac{2}{a^{2}}  \nonumber \\
\dot{\phi}^{2}(t) &=& 2{\Bigl(}\frac{1}{a^{2}}-\dot{H}{\Bigr)}.
\label{NaturalPoteqns}
\end{eqnarray}

Following \cite{EM} consider the scale factor
\begin{equation}
a(t) = A+ B\exp(ht)\,,  \label{a(t)}
\end{equation}
where $A,B,h$ are all positive constants. This universe is past
asymptotic to an Einstein static phase, since $a(t)\rightarrow A$
as $ t\rightarrow-\infty$. Thus, $A$ is identified with the radius
$a_{0}$ of the Einstein static universe. At late times, on the
other hand, $a(t)\rightarrow B\exp(ht)$ and the model approaches a
de Sitter expansion phase. The second of
eqs.~(\ref{NaturalPoteqns}) then gives
\begin{eqnarray}
\phi(b) = \sqrt{2}\int\,\frac{db}{hb}\sqrt{\frac{1-h^{2}a_{0}b} {
(a_{0}+b)^{2}}},
\end{eqnarray}
where $b:=B\exp(ht)$. The integral is easily evaluated to give
\begin{eqnarray}
\phi(b) &=& \frac{\sqrt{2}}{ha_{0}}\ln{\Biggl[}{\Bigl(}
\frac{\sqrt{ 1+h^{2}a_{0}^{2}} + \sqrt{1-h^{2}a_{0}b}}
{\sqrt{1+h^{2}a_{0}^{2}} - \sqrt{
1-h^{2}a_{0}b}}{\Bigr)}^{\sqrt{1 + h^{2}a_{0}^{2}}}\cdot  \nonumber \\
&&\cdot{\Bigl(} \frac{1 - \sqrt{1-h^{2}a_{0}b}} {1 +
\sqrt{1-h^{2}a_{0}b}}{ \Bigr)}{\Biggr]}\,,  \label{PhiExpression}
\end{eqnarray}
and with the $b$ parameterization the potential is written
\begin{eqnarray}
V(b) = \frac{3(hb)^{2} + h^{2}a_{0}b + 2}{(a_{0} + b)^{2}}\,.
\label{VExpression}
\end{eqnarray}
At this stage, an expression for $V(\phi)$ is, in principle,
obtained by inverting eq.~(\ref{PhiExpression}) and substituting
$b(\phi)$ into eq.(\ref{VExpression}). However the above form of
$\phi(b)$ precludes such a simple treatment. Nevertheless, the
general shape of $V(\phi)$ is easily determined by a parametric
plot. This is given in Fig.{\ref{ExactPotentialPic}}

\acknowledgments G.F.R.E thanks the NRF(South Africa) for funding this work, 
J.M. is supported by a research associateship of
the University of Cape Town and the Lindbury trust and C.T. acknowledges financial
support from a Sida/NRF fund and a DAA bursary. We thank Bill Stoeger for 
very helpful contributions and have enjoyed very fruitful discussions with John
Barrow, Peter Dunsby, Roy Maartens, Jean-Philippe Uzan and Amanda Weltman at 
various stages of this work.


\end{document}